# Cultural evolution shaped human's uniquely sensitive disgust avoidance

**Authors**

Edwin S. Dalmaijer [1], & Thomas Armstrong [2]

**Affiliations**

[1] School of Psychology and Neuroscience, University of Bristol, 12a Priory Road, Bristol, BS8 1TU, United Kingdom.

[2] Department of Psychology, Whitman College, 345 Boyer Ave, Walla Walla, WA, 99362, USA.

**Corresponding author**

Dr Edwin Dalmaijer, School of Psychology and Neuroscience, University of Bristol, 12a Priory Road, Bristol, BS8 1TU, United Kingdom. edwin.dalmaijer@bristol.ac.uk

**ORCID**

Dalmaijer: 0000-0003-3241-0760

Armstrong: 0000-0002-9153-8070

**Competing interests**

The authors declare that they have no competing interests (financial or other) that could have influenced or appeared to influence the work reported here.

**Acknowledgements**

We thank Dr Camilla Nord for providing feedback on an earlier version of this manuscript; and to Dr Aaron Lukaszewski, Dr Damian Murray, and Dr Joshua Tybur for providing feedback on a preprint (v2) for this manuscript in public discourse on Twitter. We are similarly grateful to anonymous reviewers who offered exceedingly useful and constructive feedback on a later version of this manuscript (preprint v4). Finally, to avoid our public embarrassment, we would like to politely request readers do not look up the publication years of the preprint and published versions of this manuscript.

**Author contributions**

ESD conceptualised the study, developed the methodology and software, conducted statistical analyses, visualised the data, and drafted and edited the manuscript. TA assisted in conceptualising the study, and reviewed and edited the manuscript.

## Abstract

Disgust is an emotion that serves to avoid contaminants, and is central to the *behavioural immune system*. While disgust-motivated avoidance occurs in bonobos and chimpanzees, humans show uniquely high levels of contamination sensitivity. There is some disagreement over to what extent human disgust avoidance is a genetic adaptation or shaped through social transmission, and whether parental modelling plays a role in the latter. Here, we address this debate by simulating 100000 years of cultural and biological evolution in various scenarios. We modelled disgust as a trait that governed the extent to which individuals forewent potentially contaminated nutrition. This lowered their risk of gastrointestinal illness, at the cost of increasing starvation risk and birth-interval. As environmental levels and costs of contamination rose, so did relative fitness of high disgust-avoidant individuals. Evolutionary shifts in disgust as polygenic trait occurred as a consequence of germ-cell mutations, but were most prominent in populations with high initial genetic variance. Crucially, cultural transmission between generations operated at a higher rate, even if parental modelling was eliminated. These results underscore the plausibility of cultural evolution's role in shaping human's uniquely sensitive disgust avoidance, and serves as an illustration of emerging theories that paint affective and cognitive mechanisms as socially transmitted rather than genetic adaptations.

## Introduction

Disgust is a basic emotion (Ekman & Friesen, 1971) evoked by "something revolting" through taste, smell, touch, or sight (Darwin, 1872). Because of its role in avoiding potential contaminants (Angyal, 1941; Curtis & Biran, 2001; Oaten et al., 2009; Rozin & Fallon, 1987; Stevenson et al., 2019), disgust is an important part of the *behavioural immune system* (Curtis et al., 2011; Schaller & Duncan, 2007). This system promotes disease avoidance not only by distancing from bodily effluvia (Strohminger, 2014), but also through more complex behavioural patterns in response to e.g. the threat of COVID-19 (Makhanova & Shepherd, 2020). Crucially, disgust helps to prevent oral incorporation of potential contaminants (Angyal, 1941; Curtis & Biran, 2001).

Non-human animals also show disease avoidance behaviours (Gibson & Amoroso, 2022; Hart, 1990; Loehle, 1995), but none as strict as humans. While our closest living relatives (Prüfer et al., 2012) show some disgust-motivated avoidance (Sarabian et al., 2017, 2018), 30-50% of bonobos still feed on banana slices placed upon or adjacent to faeces (Sarabian et al., 2018), and around 50% of chimpanzees consume food off of faecal replicas (Sarabian et al., 2017). By contrast, humans are reluctant to interact even with known replicas (e.g. fudge moulded as dog faeces) (Rozin et al., 1986). While originally observed in industrialised populations, similar contamination sensitivity is found in hunter-gatherers (Apicella et al., 2018). This much stronger avoidance of potentially contaminated stimuli suggests that high levels of disgust avoidance are uniquely human (Rozin & Fallon, 1987).

Contemporary theories generally agree that this uniquely sensitive sense of disgust developed in humans through pressures that acted on both biological and cultural evolution (Curtis et al., 2011; Curtis & Biran, 2001; Rozin et al., 2008; Rozin & Fallon, 1987; Tybur et al., 2013). However, there is notable variation in the details of these theories. Different weightings are ascribed to transmission mechanisms like genetic inheritance, parental modelling, and oblique social transmission.

In sum, a core function of disgust is to avoid potentially contaminated food. This sense of avoidance is present in many animals, but uniquely strong in humans. Here we focus on the relative contributions of biological (through genetics) and cultural (through parental or oblique social transmission) evolution. We do so for the last 100,000 years, to align with the appearance and population expansion of modern humans (Harpending et al., 1998; Rogers & Harpending, 1992).

### Evidence against parental modelling of disgust

Social transmission from parents to their offspring is central in some accounts of pathogen disgust sensitivity and avoidance (Rozin et al., 2008). Evidence for disgust socialisation comes from its surprising absence in early childhood (Rottman, 2014; Rozin & Fallon, 1987; Stevenson et al., 2010), resemblance between family members (Rozin et al., 1984; Stevenson et al., 2010), and maternal moderation of children's disgust (Olatunji & Viar-Paxton, 2020). However, some have noted that these findings could be reconciled with a genetic account (Curtis & Biran, 2001; Tybur et al., 2018).

Some have argued that the parental modelling of disgust sensitivity is supported by "little evidence" and that "much evidence contradicts [it]" (Tybur et al., 2018), and that "parental modelling is an unlikely candidate for such environmental inputs" (Tybur, 2021). Instead, the idea is that if social transmission occurs, it would mostly be oblique or horizontal (Tybur, 2021). Evidence for this view comes from several twin studies. In short, using samples of monozygotic and dizygotic twins under the assumption that they differ in genetic similarity (roughly 100% versus an average of 50%) but not in shared environment (e.g. life at home), these studies attributed half of the variance in disgust sensitivity to genetic heritability and half to non-shared environmental influences (Olatunji et al., 2019; Sherlock et al., 2016; Tybur et al., 2020). These findings did not mirror an earlier twin study that did ascribe substantial variance in disgust sensitivity to shared environmental influences (Rozin & Millman, 1987), but this has rightfully been criticised for being substantially underpowered (Tybur et al., 2018).

In sum, if parental modelling occurs in the shared environment between siblings, the above could be read as evidence against its existence.

**Do twin studies alone offer sufficient in the context of disgust avoidance?**

While useful in many contexts, twin studies are an imperfect tool. Their results are predicated upon the assumptions that siblings share the same environment, and that environment and heritability are unrelated. This assumption need not hold for disgust avoidance: children with a genetic predisposition away from disgust elicitors are less likely to be corrected by their parents. Conversely, children who do attempt to touch (or, worse, orally explore) disgust elicitors are highly likely to receive parental correction. This would be a reactive (also described as "evocative") gene-environment correlation that could inflate estimates of heritability and non-shared environmental influences (Plomin et al., 1977; Stenberg, 2013).

Attempts to empirically assess the equal environment assumption have yielded mixed results (Beckwith & Morris, 2008; Joseph, 2002; Littvay, 2012; Stenberg, 2013). For example, monozygotic twins who were misperceived to be dizygotic can be as similar as correctly identified monozygotic twins (Conley et al., 2013; Kendler et al., 1993). In addition, physical similarly between twins has been used as proxy for environmental similarity under the assumption that individuals who look alike are treated alike, and was shown not to impact the concordance in diagnosis of some (but not all) psychiatric conditions (Hettema et al., 1995).

Others have attempted to quantify genetic influences on traits through non-twin methodology, and have come to similar heritability assessments for height (Visscher et al., 2006). The same can not always be said for behavioural, cognitive, and psychological traits: twin studies overestimate their heritability in comparison to the polygenic effects captured by genome-wide association studies (Abdellaoui & Verweij, 2021). This could be partially explained by methodological difficulties in e.g. GWAS sensitivity to small and rare effects, but also by the violations of the equal environments assumption (Felson, 2014).

Simulation studies allow for the manipulation of ground-truth correlation between siblings' environments. Twin study results hold when this correlation is 1 (i.e. equal environments), but are substantially distorted when it is not (Derks et al., 2006); even to the point where a trait with equal heritability (A=44) and shared environmental influence (C=44) results in a biased estimation of primarily heritability (A=78) and almost no shared environmental impact (C=11). Systematically introducing gene-environment confounding in simulations yields similar results, and suggests that heritability is over-estimated even at low levels of confounding (Dalmaijer, 2020). Finally, heritability is over-estimated by traditional twin methodology when additive and dominant genetic effects were present for a trait that was also impacted by environments shared between siblings and between siblings and their parents (Keller et al., 2010), although the latter could additionally increase shared-environmental effects too. While more sensitive statistical approaches exist, they require extended twin family designs in which data on parents and offspring (and sometimes mates) of twins are collected (Keller et al., 2010).

Methodological work that builds on previous attempts at quantifying equal environments (Visscher et al., 2006) offers a more precise model for twin decomposition, but its authors note that this come with substantial power requirements (N>50,000 dizygotic twin pairs) and with different but equally strong assumptions (Hwang et al., 2021). Given the small scale of previous twin studies on disgust (Olatunji et al., 2019; Rozin & Millman, 1987), even those that used twin registries (Sherlock et al., 2016; Tybur et al., 2020), this was not a feasible approach for them.

In sum, while several twin studies on disgust sensitivity have been conducted, they should not necessarily be used to assess the genetic heritability of disgust sensitivity in the absence of convergent data from approaches that directly measure individuals' genetics.

## Directly testing evolutionary hypotheses

Here, we directly compared genetic and cultural transmission in agent-based simulations of 100,000 years of evolution. In our model, the trait disgust avoidance was directly associated with the intake of potentially contaminated food, which was a subset of all available nutrition in an environment (level of contamination). This was a simplification in which all potentially contaminated food was distinguishable from safe food. Ingestion of contaminated food resulted in a chance of developing gastrointestinal illness, and thus associated with mortality (cost of contamination). Nutritional intake was associated with survival and with the duration of postpartum amenorrhoea, with relative malnutrition increasing both mortality and inter-birth intervals (both reducing fitness).

We modelled disgust as a polygenic trait passed down through sexual reproduction, and as cultural trait imperfectly copied from previous generations (Cavalli-Sforza & Feldman, 1973b; Eerkens & Lipo, 2005; Hamilton & Buchanan, 2009). The resulting phenotype was an equal weighting of the genotype and the socially learned trait. In simulated environments, we systematically varied the available nutrition, and the level and cost of contamination. Parameters in our simulation were informed by research on contemporary hunter-gatherer societies (outlined comprehensively in the Methods section).

## Notable limitations on our evolutionary modelling

While limitations are generally provided in the Discussion, we felt it prudent to also contextualise the current study by noting what it does and does not include in its modelling.

First, our model does not separate between the mechanisms that detect disgust. An individual could identify a disgusting stimulus through sensory (e.g. the smell of rot) or semantic (e.g. cultural norms against consuming conspecific cadavers). What we model is disgust avoidance, i.e. the passing on food sources given knowledge about their disgusting state.

Second, our model is agnostic to how specific mechanisms are inherited. Social transmission is often described as a way to calibrate which stimuli should invoke disgust (Curtis & Biran, 2001; Tybur et al., 2013), e.g. for specific food items (Cashdan, 1994) or even arbitrary objects (Repacholi, 1998). We simply assume that a form of disgust avoidance exists and is transmitted.

Third, our model independently modelled genetic and socially transmitted traits, but no explicit interaction between the two. One benefit is that we could model both within the same simulations, which was computationally efficient. It also mapped onto the notion that all psychological traits are a joint product of genes and environment (Cosmides & Tooby, 1997).

Fourth, we did not distinguish function between the cultural and genetic traits. Both operated on the same inputs and outputs, and were equally important in shaping behaviour. This could have been more nuanced, but any attempt at modelling this would have been highly speculative as it remains unclear what the exact mappings are between genotype, socially transmitted traits, and disgust sensitivity and avoidance are.

In sum, our model treats disgust avoidance as a singular phenotype comprised of an independent genotype and socially transmitted trait. However, this does not mean it is a black box on which any label could fit. The avoidance of disgusting stimuli is a specific behaviour well characterised in human adults (Armstrong et al., 2014; Dalmaijer et al., 2021; Rozin et al., 1986) and children over the age of 5 years (Alladin et al., 2024; Stevenson et al., 2010); and its direct impact on food rejection is obvious in humans (Apicella et al., 2018; Gumussoy et al., 2021, 2025; Rozin et al., 1986; Stern et al., 2001) and other apes (Sarabian et al., 2017, 2018). While higher disgust avoidance could result in reduced illness-related mortality, it could similarly contribute to increased malnutrition-related mortality and an increase in the duration of inter-birth intervals.

Thus, whether disgust avoidance bequeathed a fitness advantage and was passed down depended on environmental characteristics. In other words, our model is not a self-fulfilling prophecy.

## Methods

### Experimental design

We tested how culturally and genetically transmitted traits behaved under different mechanisms and in different environments. Due to the practical difficulties associated with the experimental manipulation of human evolution, we employed agent-based simulations. Each simulation comprised a combination of specific parameters that governed the initial genetic variance, the parental influences in cultural transmission, nutritional availability, environmental contamination, and the deadliness of contaminated nutrition. The effect of these parameters was assessed in reference to control conditions without selection pressures. In such control environments, there was no environmental contamination or mortality cost to contamination.

### Trait inheritance

Disgust was modelled as polygenic trait (mimicking other psychological traits, e.g. (Levey et al., 2020)), and transmitted through sexual reproduction. The genotype was initialised with 60% avoidance of contaminated nutrition (mirroring bonobo (Sarabian et al., 2018)); and a variance of $2n\mu/S$=0.0132 (Barton, 1986), with $n$=10 loci contributing equally to the trait, mutation rate per locus $\mu$=6.6e-5, and stabilising selection $S$=0.1 (de Vladar & Barton, 2014). Germ cell mutation probability was 1.1e-8 per position per haploid genome (Roach et al., 2010). Assuming ~6000 bases (size of AURKB, implicated in polygenic anxiety (Levey et al., 2020)), this translated to $1-(1-1.1e\text{-}8)^{6000}$=6.6e-5 per gene. This is likely a high estimate, because de novo mutations are not equiprobable across the genome (Genome of the Netherlands Consortium et al., 2015), so we also ran controls with lower initial variance (1.32e-3).

When germ cell mutations occurred, allele deviations were sampled from a normal distribution with $\mu$=0 and $\sigma$=0.025. Beneficial and detrimental mutations were thus equally likely, and smaller were more likely than larger fitness effects. For context, in a monogenic scenario with 11088 kJ daily nutrition, 0.1 contamination probability, 5.1114e-9 probability of death per contaminated kJ, and alleles D=1.0 and d=0.975 (=$1-\sigma_{mutation}$), selection coefficients were roughly in the middle of the typical range (Loewe & Hill, 2010): $s_{DD}$=1.62e-3, $s_{Dd}$=7.51e-4, and $s_{dd}$=0 (best relative fitness). This approximated typical distributions of mutational effects, which are wide and unpredictable (Keightley & Eyre-Walker, 2010).

Disgust was also modelled as culturally transmitted trait, using established models (Cavalli-Sforza & Feldman, 1973b; Eerkens & Lipo, 2005; Hamilton & Buchanan, 2009) in which individuals in generation *t+1* imperfectly copy trait *s* from their parents with error $\varepsilon$, or from prior generations *t* with probability $\lambda$ (Equation 1). The trait started at barely any culturally transmitted disgust avoidance, at *s(t)*=0.95 with a variance of 6.25e-4. Parameter values were based on empirical findings, with $\varepsilon$=0.025 (Eerkens, 2000; Hamilton & Buchanan, 2009) and $\lambda$=0.38 (Hamilton & Buchanan, 2009).

(1) $$s(t+1)=\lambda\,\bar{s}(t)[1+\epsilon(t)]+(1-\lambda)\,s(t)[1+\epsilon(t)]$$

**Environments**

Environmental nutrition $v$ was set at 6650, 8866, 11083, 13300, or 15516 kilo-Joule (kJ) per day; 60-140% of typical adult male intake (Pontzer et al., 2012). The level of contamination $\eta$ varied from 0 to 50% of available nutrition; and the cost of contamination $\kappa$ was implemented as independent probability of death per ingested contaminated kJ of 0, 2.0883e-9, 5.1114e-9, 8.2622e-9, 1.1551e-8, or 1.4993e-8. With an intake of 11083 kJ/day and 10% contamination, this corresponded to a probability of death-by-contamination of 0%, 2.75%, 6.6%, and 10.45%, 14.3%, and 18.15% over the average lifespan of 33 years (not counting child mortality). In reality, gastrointestinal illness and fever caused 13.6% (5.5+8.1) and 34.9% (13.2+21.7) of deaths among hunter-gatherer and settled Ache, respectively (Gurven & Kaplan, 2007).

**Nutritional intake**

Individual nutritional intake *i* (Equation 2) was the sum of consumed contaminated and uncontaminated nutrition, but never more than daily requirement $\rho_{male}$=11083 or $\rho_{female}$=7853 kJ (Pontzer et al., 2012). Uncontaminated intake (Equation 3) was governed by environmental contamination $\eta$, and daily nutritional availability $v$. Crucially, contaminated intake (Equation 4) was not only determined by $\eta$ and $v$, but also impacted by trait $\tau$: the unweighted average of genetic and cultural disgust. This trait determined the proportion of contaminated nutrition individuals consumed, so lower values indicate higher levels of disgust-motivated avoidance.

(2) $$i=min(\rho, i_{contaminated}+i_{uncontaminated})$$

(3) $$i_{uncontaminated}=(1-\eta)\,\nu$$

(4) $$i_{contaminated}=\tau\,\eta\,\nu$$

**Mortality**

Mortality (Equation 5) at age *x* was computed through the Siler hazard function (Equation 6) (Siler, 1979), disgust-related mortality (Equation 8) weighted by proportion $\gamma$ (Equation 7); and nutrition-related mortality (Equation 9).

(5) $$m(x,i)=\gamma f(x,i)+(1-\gamma)h(x)+n(i)$$

Siler parameters were based on those found in Hadza, Ache, Hiwi, !Kung, and Agta (Gurven & Kaplan, 2007): initial infant mortality $\alpha_1$=0.422, infant mortality decline $\beta_1$=1.131, general mortality $\alpha_2$=0.013, initial adult mortality $\alpha_3$=1.47e-4, and adult mortality increase $\beta_3$=0.086.

(6) $$h(x)=\alpha_1 \exp(-\beta_1 x)+\alpha_2+\alpha_3 \exp(\beta_3 x)$$

The proportion of mortality affected by disgust is computed from cost of contamination $\kappa$, environmental contamination $\eta$, and the lowest of daily nutrition $\nu$ or male requirement $\rho_{male}$.

(7) $$\gamma=1-(1-\kappa)^{(365.25\,\eta\, min(\nu,\rho_{male}))}$$

Disgust-associated mortality was directly impacted by disgust trait $\tau$. Its gradual emergence over childhood was normally distributed with $\mu$=5 and $\sigma$=3; because children in industrialised nations develop disgust between ages 3 and 5 years (Rozin & Fallon, 1987; Stevenson et al., 2010), and because children in hunter-gatherer societies function as "helpers at the nest" (Hames & Draper, 2004), with Hadza children engaging in food acquisition and processing from as early as 3.5 years (Hawkes et al., 1995).

(8) $$f(x,i)=\Phi\left(\frac{x-\mu}{\sigma}\right)\left(1-\left(1-\kappa^{365.25\,\eta\, i\,\tau}\right)\right)$$

Nutrition-based mortality was computed with nutrition-specific $\alpha_2$, set at -0.001 for male and -0.001($\rho_{male}/\rho_{female}$) for female individuals.

(9) $$n(i)=\exp(\alpha_{2,nutrition}\, i)$$

**Fertility**

While the relationship between dietary restrictions and female fertility is complex, under-nutrition increases birth intervals in hunter-gatherers (Spielmann, 1989) and forager-horticulturalists (Tracer, 1996). Thus, in addition to mortality, nutritional intake impacted postpartum amenorrhoea in our simulations. The yearly cumulative probability of menstruation recovery was 0, 0.49, 0.59, 0.72, 0.86, and 1 in mothers with sufficient nutrition, and weighted down by insufficient nutrition in years 2 and 3 (Tracer, 1996).

Fertility onset was normally distributed with $\mu$=16.0 and $\sigma$=2.5 for female, and $\mu$=19.5 and $\sigma$=3.4 for male individuals (based on Savannah Pumé, (Kramer et al., 2017)); and offset with $\mu$=45.0 and $\sigma$=2.5 for female (Austad, 1994; Vinicius et al., 2014), and $\mu$=55.0 and $\sigma$=3.4 for male individuals (based on !Kung and Hadza, (Vinicius et al., 2014)). Fertile individuals were randomly paired, reflecting high marriage rates among e.g. !Kung (Marshall, 1959). We did not model polygeny, because of its low prevalence in many hunter-gatherer societies (Marshall, 1959; Vinicius et al., 2014; Walker et al., 2011).

The probability of a baby being born to fertile pairs was initially set at 0.42, and allowed to vary between 0.32 and 0.52 with yearly steps of 0.002 (up if population size decreased, down if it increased or surpassed 5000). This resulted in about 7 children per mother at completed fertility, mirroring Savannah Pumé (Kramer et al., 2017). Sex ratio at birth was kept at 1.076 (based on Savannah Pumé, (Kramer et al., 2017)); slightly more skewed towards male than in industrialised nations, where birth sex ratios range from 1.031 to 1.067 (Chao et al., 2019).

### Statistical Analysis

Each unique combination of simulation parameters was applied in 10 independent runs. In each run, populations were initialised with N=1000, and allowed to grow to N≈5000. Averages and pooled standard deviations were computed over all 10 runs within each simulation, or the subset of runs that did not end in population collapse. Statistical tests were not employed, as they would result in trivial significance due to the high number of total individuals in each simulation (N≈50000). Effect sizes for evolutionary shifts in traits were computed as changes in disgust-approach proportions compared to chance level drift. Chance level was defined as the average change in traits in control conditions without contamination ($\eta$=0) and with no cost to contamination ($\kappa$=0).

### Data and code availability

Simulation code and data is available from a public GitHub repository: https://github.com/esdalmaijer/2020_disgust_evolution.

## Results

### Disgust is selected for in a typical environment

Over evolutionary time, in an environment with sufficient pressure (environmental nutrition $\nu$=11083 kJ/day, level of contamination $\eta$=0.1, cost of contamination $\kappa$=1.1551e-8), selection occurred for disgust avoidance (Figure 1A). In a control environment without pressure ($\nu$=11083 kJ/day, $\eta$=0, $\kappa$=0), lower drift occurred (Figure 1B). Control environments with $\eta$=0 and $\kappa$=0 were used to empirically establish “chance level” drift for all scenarios and nutrition levels.

Evolutionary shifts exceeded chance levels to a greater extent as the cost of contamination ($\kappa$) increased (Figure 1C), indicating generations were increasingly less likely to consume contaminated nutrition. For the polygenic trait, change was driven by an increase in more disgust-avoidant, and a decrease of less disgust-avoidant alleles in the genotype.

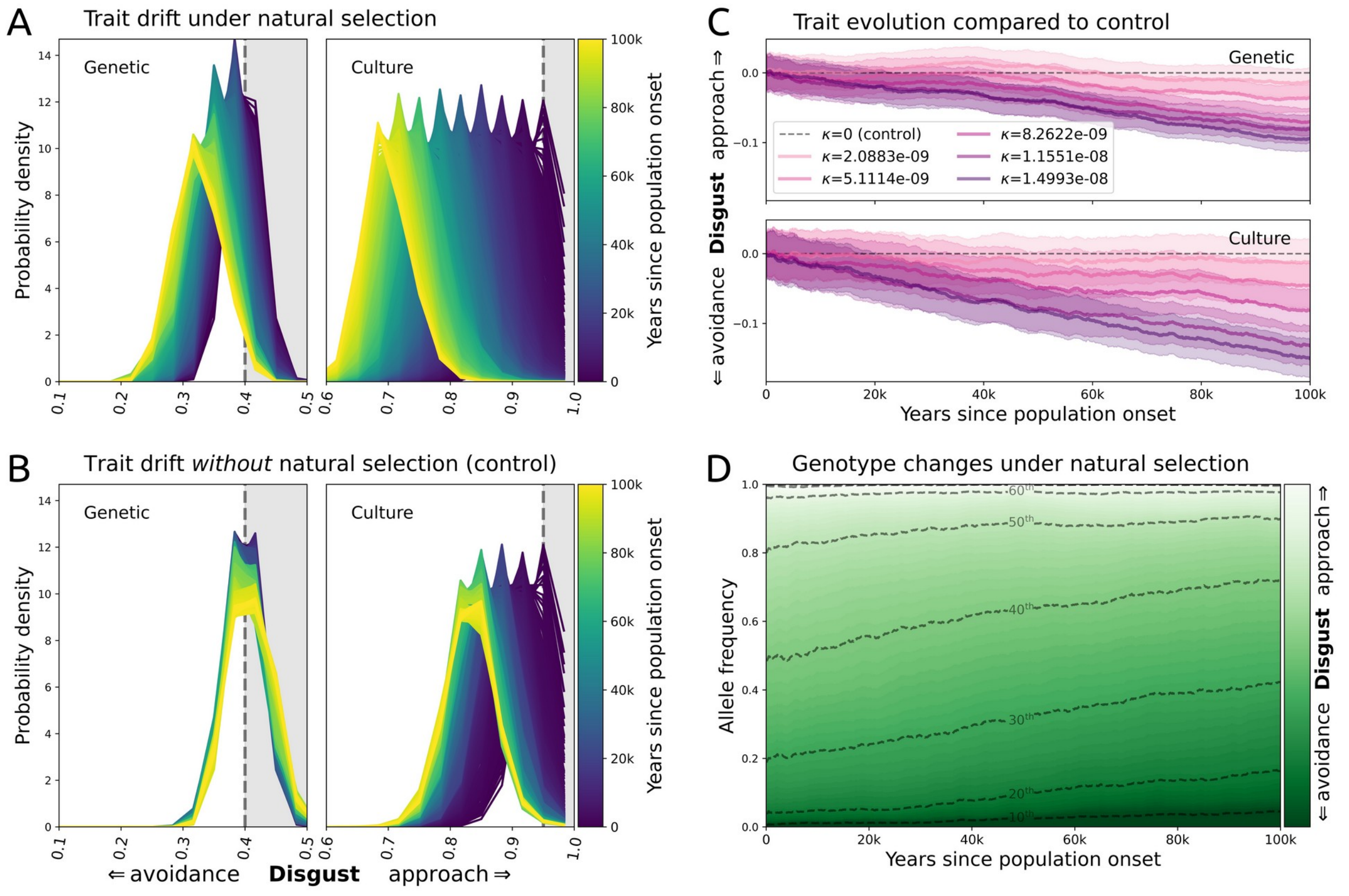

***Fig. 1*** *– Results of 100,000 year simulations of disgust, a core trait in the human behavioural immune system. Higher disgust-motivated avoidance reduced intake of potential contaminants, and thus reduced mortality due to gastrointestinal illness; at the cost of reducing total nutritional intake, and thus increased birth interval and mortality due to malnourishment.* ***A)*** *Probability density of the polygenic and cultural trait components in an environment with selective pressure (ν=11083 kJ/day, η=0.1, κ=1.1551e-8), each averaged across 10 simulation runs (brighter colours reflect later years). The distribution moves towards better fitness at more disgust-motivated avoidance, at higher magnitude for the cultural compared to the genetic trait.* ***B)*** *Probability density for the trait components in a control simulation (ν=11083 kJ/day, η=0, κ=0), averaged across 10 simulation runs each. While modelled as a Markov process with no inherent direction, culture drifted towards improved disgust-motivated avoidance, likely because its starting point near the maximum value prevented upward drift.* ***C)*** *Averages (line) and pooled standard deviations (shading) over 10 simulation runs each for traits passed down through genetic (top) or cultural transmission (bottom) under varying cost of contamination (κ).* ***D)*** *Changes in allele frequency over time in a population with a high initial genetic variance. Alleles that contribute to increasing disgust-motivated avoidance became more prevalent.*

**Cultural evolution outpaces biological evolution for disgust**

An overview of disgust evolution under different levels of nutritional availability and contamination is provided in Figure 2. The main pattern is that selection for disgust-motivated avoidance of contaminated nutrition is stronger as the levels of environmental contamination and the cost of contamination increases. While both the cultural (Figure 2A) and the polygenic (Figure 2B) disgust components showed adaptation, the magnitude of this shift exceeded chance levels to a greater extent for the culturally compared to the genetically transmitted trait (Figure 2C).

These results suggest that the degree of selective pressure for disgust increased with contamination danger, and that cultural evolution of disgust operated at a higher rate.

**Disgust is maladaptive in low-nutrition environments with many low-risk contaminants**

While higher levels of disgust were selected for in most environments, there were notable exceptions to this rule. Specifically, the disgust-motivated avoidance of contaminated nutrition were maladaptive in at least two environments. These were characterised by relatively low nutritional availability ($v$=8866 kJ/day, 80% of average male intake), combined with a high level of environmental contamination ($\eta$=0.5) but low cost of contamination ($\kappa$=0 or $\kappa$=2.0883e-9). In these environments, an evolutionary shift towards less disgust-motivated avoidance occurred for the polygenic trait. (It did not occur for the cultural trait, as this was initialised near ceiling.)

These results indicate that disgust was maladaptive in environments where food was scarce and likely to be contaminated, but not likely to be lethal.

**Initial genetic variance determines rate of biological evolution**

The increase in disgust-avoidant genotypes was driven by better fitness of individuals already in the more contamination-avoidant tail at population initialisation. Populations initialised with high initial variance (Figure 2C) showed greater genetic adaptation than populations with lower initial variance (Figure 2D). These results indicate that when genetic evolution was dependent on mutations, cultural evolution rapidly outpaced it.

**Eliminating parental modelling slows down cultural evolution**

Additional controls were run to simulate the (unlikely) scenario of absent parental modelling by setting conformist bias parameter $\lambda$=1. When initialised with high genetic variance, changes (compared to random drift) in culturally and genetically transmitted traits were almost equally matched (Figure 2E), but not when initial genetic variance was less high (Figure 2F).

These results indicate that the rate of cultural evolution was stunted when parents did not directly transmit disgust to their offspring, but the mechanism still drove change more quickly than mutation-driven genetic adaptation.

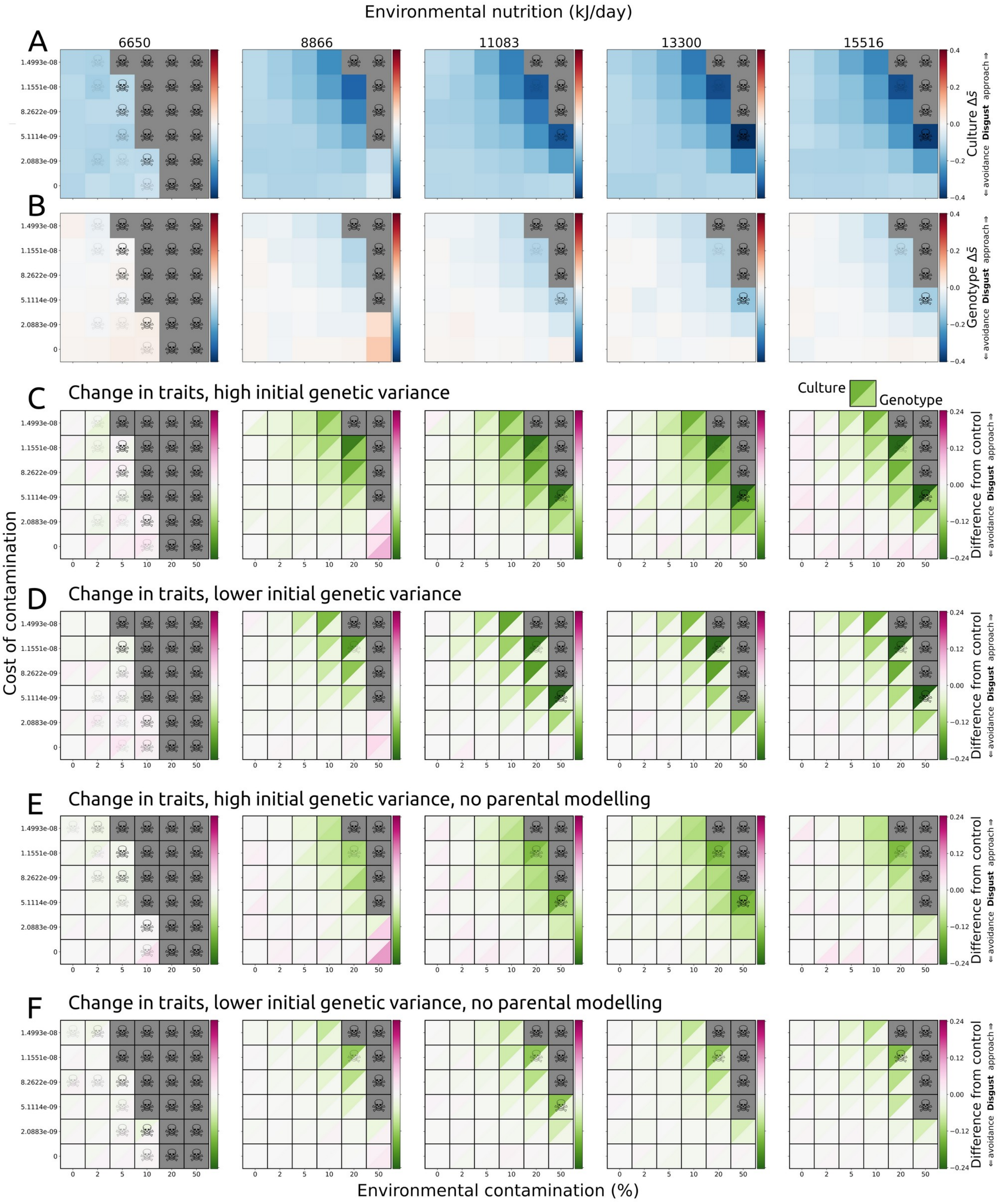

Environmental nutrition (kJ/day)
A
6650
8866
11083
13300
15516
Culture Δs̄
⇐ avoidance Disgust approach ⇒
B
Genotype Δs̄
C
Change in traits, high initial genetic variance
Culture
Genotype
Difference from control
D
Change in traits, lower initial genetic variance
E
Change in traits, high initial genetic variance, no parental modelling
F
Change in traits, lower initial genetic variance, no parental modelling
Cost of contamination
Environmental contamination (%)

***Fig. 2** (previous page) – Results of all simulated environments, with each square representing an average of 10 simulation runs. Each block represents environments with different nutritional availability (in kilo-Joule per day). Within each block, the proportion of contaminated nutrition (η) increases over the x-axis, and the probability of death per contaminated kJ (κ) increases along the y-axis. The origin (η=0, κ=0) is a control condition, as it presents no contamination danger, and thus no selection pressure. Skulls indicate total population collapse, with higher opacity indicating more runs ended in collapse, and grey squares indicating no populations survived. **A)** Total shift in average cultural trait over 100,000 years. Shifts are always towards more disgust-motivated avoidance, even in the control condition, because the trait's starting point is to approach all the contaminated nutrition. **B)** Total shift in average polygenic trait over 100,000 years. **C-F)** Magnitudes of evolutionary shifts compared to their control condition (no selection pressure at η=0, κ=0), split between cultural (upper-left triangles) and genetic (lower-right triangles) inheritance. Simulated scenarios implemented high (C and E) or less high (D and F) initial genetic variance, and did (C and D) or did not (E and F) implement parental modelling.*

## Discussion

Inspired by prominent theories of the behavioural immune system's development, we simulated 100,000 years of human life. We focussed on disgust because of its central role in pathogen avoidance (Angyal, 1941; Curtis et al., 2011; Rozin & Fallon, 1987; Schaller & Duncan, 2007; Stevenson et al., 2019) and its uniquely prominent profile in humans (Apicella et al., 2018; Rozin et al., 1986; Rozin & Fallon, 1987). Our results showed that disgust avoidance was selected for in environments with sufficiently available nutrition, unless food contamination was abundant and of low pathogen risk. Furthermore, the socially transmitted trait shifted further than its polygenic counterpart in the same amount of time, suggesting that cultural evolution operated more quickly. This was impacted by parental modelling, as its removal reduced the pace of cultural evolution. Finally, the rate of genetic adaptation was particularly slow when a population's initial genetic variance was low, and adaptation thus relied more strongly on rare mutations.

Our findings broadly align with existing work on the relative paces of biological and cultural evolution, which suggests that genetic transmission is outpaced by social transmission in the long term (Perreault, 2012). What we did not replicate in our study is an accelerating pace of cultural evolution, perhaps as a consequence of our choice to implement a relatively uncomplicated model (Cavalli-Sforza & Feldman, 1973b). Instead, we found the rate of adaptation to be constant, albeit still proportional to environmental pressures (see Figure 1C).

More specific to disgust, our work aligns with theories that propose human disgust sensitivity emerged through selective pressures on both genes and learned behaviours (Curtis et al., 2011; Rozin et al., 2008; Rozin & Fallon, 1987; Tybur et al., 2013). Where our findings differ is in support for the idea that parents do not socially transfer the trait to their offspring (Tybur, 2021; Tybur et al., 2018). When parental modelling was removed as a mechanism of inheritance, the pace of cultural evolution was hampered to the point of parity with biological evolution or below (contrast Figures 2C and 2E), provided initial genetic variance was high. It is tempting to consider this rough equality between the two as analogous to that between estimates of the influence of genetic heritability and non-shared environmental influences from twin studies (Olatunji et al., 2019; Sherlock et al., 2016; Tybur et al., 2020). Nevertheless, these are not directly comparable estimates. Our results suggest that, even in the presence of oblique transmission, parental modelling is a perfectly plausible mechanism for inter-generational transfer. The fact that its presence substantially improved adaptability disagrees with the suggestion that parental influences might be diluted by those from other individuals during children's development (Tybur, 2021).

As highlighted in the introduction, avoidance of potential contaminants is much stronger in humans (Apicella et al., 2018; Rozin et al., 1986) compared to bonobos (Sarabian et al., 2018) or chimpanzees (Sarabian et al., 2017). Thus, while disgust-motivated avoidance is not unique to humans, the high degree to which humans display it is. This is mirrored in cultural evolution: while some have suggested it is uniquely human (Tennie et al., 2009), plenty of evidence suggest that it is not. The type of cumulative culture that relies on form-copying social learning (as it is modelled here) occurs in non-human primates too (Boesch et al., 2020; Claidière et al., 2014; Matsuzawa et al., 2008; Musgrave et al., 2020; Schofield et al., 2018; Vale et al., 2017), and cannot be explained through (re)innovation (Koops et al., 2022). As with disgust itself, it may not be the existence of cumulative culture that is unique to humans, but rather the degree to which it is supported by a range of social and cognitive skills (Migliano & Vinicius, 2022; Whiten, 2021). Hence, it is possible that humans developed their uniquely strong propensity for disgust due to their uniquely strong ability to socially transmit traits between generations.

**Limitations**

Disgust is theorised to serve other adaptive functions not addressed in our model, including mate selection and coordinating moral condemnation (Tybur et al., 2013). However, most theories view pathogen or "core" disgust as the original form, which was then co-opted to serve other functions (Rozin et al., 2008; Rozin & Fallon, 1987; Tybur et al., 2013). In addition, whether these functions build on core disgust, or only share superficial overlap grounded in common parlance (Fiske, 2020), remains an open question (Armstrong et al., 2021).

Because of the above, our model did not incorporate assortative mating. This, and other types of dependencies between genotype and phenotype through cultural expressions, have been the focus of previous population-level modelling efforts (Cavalli-Sforza & Feldman, 1973a). We modelled genetic and cultural inheritance as two independent underlying processes that acted on the same phenotype, which is defensible as the most conservative approach in the absence of strong data on potential sources of gene-culture coevolution in the context of disgust. However, future work could incorporate different types of coevolution scenarios to evaluate their relative likelihood.

Another simplification in our model was in the lack of niche construction. We did not incorporate in the idea that type and probability of cultural transmission could differ as a function of individuals' cultural background (Ihara & Feldman, 2004), or the population's demographic structure (Fogarty et al., 2013, 2019).

## Conclusion

Our agent-based simulations suggest that under selective pressure of pathogens, cultural evolution outpaced genetics in shaping human's unique propensity for disgust avoidance. Our results support the hypothesis that this element of the behavioural immune system evolved, at least in part, as a "cognitive gadget" (Heyes, 2020): a mental mechanism passed down generations through social transmission.

**Preprint version history**

*ArXiv v5, 2026-10-09*

- **Precision in scholarship**. Anonymous reviewers of this manuscript greatly help to make our reflection of the theoretical background to this paper much less coarse. Several arguments put forward in past versions were too imprecise, general, and under-explained. This includes reflections of others' work and theories, which were captured only in the highlights that we read in them. Additional nuance was crucial to not make caricatures out of researchers' positions, and the current introduction is an attempt to more accurately reflect cited work (e.g. using quotes), but without being overly expansive in our review of the literature.
- **Scholarship of twin studies**. One specific area of contention was how much evidence twin studies offer on the heritability of disgust sensitivity. We contended it was insufficient to rule out parental modelling, and were to curt in our treatment of twin studies in the opinion of several readers. This was not an entirely unfair criticism, and it was offered constructively, so we expanded upon our discussion. In short, we still make the same point, but in a more fair and complete overview of the benefits and limitations of twin study designs. (On a more editorial note, this was a tough decision, as the treatise on twin studies dramatically expands the introduction on what is almost a tangent. Readers would be better served with a briefer introduction, but as the twin studies are core to arguments about a potential absence of parental modelling, it wouldn't serve to not touch upon them.)
- **Head-on addressing of modelling limitations**. There were some misgivings and misunderstandings of our model, and what it did or did not reflect. The Introduction section now has a heading that directly addresses this, and further clarity should come from the narrowed scope of our interpretation of our findings. These are now firmly confined to human disgust avoidance. (This is part of the improved precision in scholarship, as previously our writing was more expansive, and could have easily be taken for an attempt to model the entire behavioural immune system. We didn't necessarily intend for this, but the writing was on the slightly grandiose side, and readers would have had to read the Methods section to confirm it was disgust avoidance that we were modelling specifically.)

*ArXiv v4, 2022-02-14*

- **Limitations**. Improved Discussion section on limitations, which now includes the aspects of disgust that remain unaddressed in our model (e.g. sexual and moral disgust). Additionally, the section addresses evolutionary processes that were not modelled, including gene-culture coevolution and demographic factors that could affect social transmission.
- **Methodological overview**. The Methods section now opens with an overview of the study, and includes a new section on the (lack of) statistical testing.

*ArXiv v3, 2021-12-16*

- **Scholarship**. An anonymous reviewer of this manuscript pointed out that the model of cultural evolution attributed to Eerkens & Lipo (2005) is highly similar to that proposed by Cavalli-Sforza & Feldman (1973). We now appropriately cite them.
- **Nutrition-illness trade-off**. In previous versions, disgust was modelled as a dampening factor on gastrointestinal illness due to its effect on the avoidance of contaminated food. The flip side of this, missed nutrition, was not incorporated. The current model does incorporate missed nutrition, which now impacts on survival and on postpartum amenorrhoea.
- **Phylogenetic starting point**. The genetic starting point for disgust avoidance is now set at an average of 40% feeding of visibly contaminated food. This is based on work by Sarabian and colleagues (2018) who show that about 30-50% of studied bonobos ate banana slices that were atop or directly adjacent to faeces; and in chimpanzees who avoid about 50% of olfactory and tactile faecal replicas (Sarabian et al., 2017). The cultural starting point was set near 100% of feeding on contaminated food, i.e. near 0% disgust-motivated avoidance.

*ArXiv v2, 2020-09-06*

- Corrected reference numbering. (We forgot to hit Zotero's refresh button after changing the order of several sentences. Oops!)

*ArXiv v1, 2020-08-30*

Original submission.